\documentclass[11pt,a4paper]{article}
\usepackage[T1]{fontenc}
\usepackage[utf8]{inputenc}
\usepackage{authblk}
\usepackage{parskip}
\usepackage{graphicx}
\usepackage{amssymb}

\usepackage{lipsum}
\makeatletter
\renewcommand{\maketitle}{\bgroup\setlength{\parindent}{0pt}
\begin{flushleft}
  \textbf{\@title}\\*[0.3cm]
  
  \@author
\end{flushleft}\egroup
}
\makeatother

\usepackage{authblk}
\makeatletter
\renewcommand\AB@affilsepx{. \protect\Affilfont}
\makeatother

\title{\huge The language of exoplanet ranking metrics needs to change}
\author[1]{\small Elizabeth Tasker}
\author[2]{\small Joshua Tan}
\author[3]{\small Kevin Heng}
\author[4]{\small Stephen Kane}
\author[5]{\small David Spiegel}
\author[+]{\small the ELSI Origins Network Planetary Diversity Workshop}
\affil[1]{\footnotesize Institute of Space and Astronomical Science, Japan Aerospace Exploration Agency, Yoshinodai 3-1-1, Sagamihara, Kanagawa, Japan}
\affil[2]{\footnotesize Instituto de Astrofísica, Pontificia Universidad Católica de Chile, Santiago, Chile}
\affil[3]{\footnotesize University of Bern, Center for Space and Habitability, Sidlerstrasse 5, CH-3012, Bern, Switzerland}
\affil[4]{\footnotesize Department of Physics \& Astronomy, San Francisco State University, 1600 Holloway Avenue, San Francisco, CA 94132, USA}
\affil[5]{\footnotesize Analytics \& Algorithms, Stitch Fix, San Francisco, CA 94103, USA}
\affil[-]{\footnotesize \newline\newline {\bf $^{+}$EON Workshop Members}: Ramon Brasser (Earth Life Science Institute, Tokyo Institute of Technology), Andrew Casey (Institute of Astronomy, University of Cambridge), Steven Desch (School of Earth and Space Exploration, Arizona State University), Caroline Dorn (Space Research \& Planetary Sciences, University of Bern), Christine Houser (Earth Life Science Institute, Tokyo Institute of Technology), John Hernlund (Earth Life Science Institute, Tokyo Institute of Technology), Marine Lasbleis (Earth Life Science Institute, Tokyo Institute of Technology), Matthieu Laneuville (Earth Life Science Institute, Tokyo Institute of Technology), Anne-Sophie Libert (naXys, University of Namur), Lena Noack (Department of Reference Systems and Planetology, Royal Observatory of Belgium), Cayman Unterborn (School of Earth and Space Exploration, Arizona State University), June Wicks (Department of Geosciences, Princeton University)}

\date{}
 
\begin{document}
  \maketitle
{\it Published in Nature Astronomy, Comment, 2nd February 2017}  
\vspace{0.5cm}

{\bf We have found many Earth-sized worlds but we have no way of determining if their surfaces are Earth-like. This makes it impossible to quantitatively compare habitability, and pretending we can risks damaging the field.}
\vspace{0.5cm}

Over 3,000 planets have been confirmed beyond our Solar System, with approximately a third smaller than twice the radius of Earth \cite{stats}. This swarm of approximately Earth-sized worlds has led to intense speculation about whether such planets might also harbour life. 

In the next decade, telescopes capable of tackling this question will start to be available. But with thousands of planets and observational hours in high demand, target prioritization is essential.  This has led to the development of metrics to rank planets for future habitability studies.  Three of the most commonly used metric are described below; all of them attempt to identify the exoplanets most likely to show signs of life.

Unfortunately, these metrics have a dark side. Their significance is frequently misinterpreted by the media, and occasionally within the scientific community, as being a quantitative measure of planet habitability. Such a measure is currently impossible. The properties we can observe are not directly connected to habitability and our single reference point for a habitable planet (you're standing on it) restricts our understanding of the dependent factors.

This misuse of target selection metrics has also been intentionally promoted in a desire to publicise scientific results. This is effectively pseudo-science and the consequences are serious. Poor understanding of the metric outputs risks wasting resources on targets unlikely to show biological activity, exhausting public interest and damaging the respectability of exoplanet studies. Unless tackled by the research community, the result will be a global reluctance to fund future projects. To search for a way to improve this situation, we need to consider what planetary environments we aim to detect, exactly what properties can be measured and how we might communicate this accurately to a wide audience.

\subsubsection*{Detectability, not habitability}

The term `habitable' is generally understood to refer to an environment that could support any form of life. In practice, this definition is unhelpful since extraterrestrial life is only scientifically valuable if it can be detected. This restricts us to biological activity that creates a distinctive change in the composition of a planet's atmosphere, or in the wavelengths of radiation reflected from the planet surface \cite{biosig}. A subsurface ocean like that on Europa might host an ecosystem, but unless there is substantial exchange of organic material with the surface then it will remain undiscovered on worlds too distant for robotic exploration. Similarly, a planet too distant for spectroscopic observations is uninteresting for the detection of life.

Since successful detection also requires us to recognise any biosignatures, plans for finding habitable planets have focussed on Earth-like life. More specifically, this has meant searching for worlds that can support liquid water on the surface.

The conditions relevant for detectable habitability are therefore those on the planet's surface. Unfortunately, observing the surface is challenging even for the most ambitious future missions and may be perpetually blocked from direct view by the planet's atmosphere. We are therefore forced to estimate surface conditions based on what properties we can observe. For the majority of exoplanet discoveries, this comes down to only two independent measurements: the incident flux from the star and the value of either the planet's radius or minimum mass. Not only is this information sparse, but its relationship to habitability is far from linear. 

\subsubsection*{What we can determine}

The incident flux from the star can be used to calculate an `equilibrium temperature' at the planet's position. This depends on the stellar luminosity, distance and (where known) the planet's orbital eccentricity and albedo. However, this is not the same as surface temperature. How the two are related depends on the planet's atmosphere.

The Earth's equilibrium temperature is 255\,K, well below the freezing point of water at 273\,K. The greenhouse gases in our atmosphere raise the surface temperature by 33\,K to bring the global average value to 288\,K. By contrast, Venus's far denser atmosphere moves the equilibrium temperature from about 300\,K (using an Earth-like albedo as commonly done in exoplanetary observations) to a lead-melting 735\,K. 

The situation is further complicated by the most observable targets for spectroscopic studies orbiting close to cool M-dwarf stars. These worlds risk tidal locking, where one side of the planet permanently faces the star. The equilibrium temperature suggests these planets would suffer from atmospheric collapse as gas condenses on the dark hemisphere. However, this can be avoided if the planet's atmosphere can redistribute the heat \cite{tidallock}. Similar considerations apply to planets on elliptical orbits with potentially crippling season changes \cite{eccentric}.

In order to translate between equilibrium and surface temperatures we can apply atmospheric models. This is in theory a rigorous approach, but unfortunately these models are time consuming and rely on a detailed knowledge of atmospheric parameters (gaseous composition, pressure and temperature profile, and so on) which are generally poorly (if at all) known. This makes them unusable as a target selection tool. Instead, using the equilibrium temperature to identify potentially habitable worlds must assume an Earth-like atmosphere.

This brings us to the problem of planet size. Just as equilibrium temperature is a poor proxy for surface temperature, an Earth-sized planet does not mean an Earth-like composition. 

The minimum requirement for an Earth-like atmosphere or any form of Earth-like life is a solid surface. This has to be deduced from either the planet's minimum mass (if the planet was detected by the radial velocity technique) or its radius (if detected when transiting the star). If the planet has Jupiter proportions, then it is safe to say there is no rocky surface. On the other hand, it is not clear if the populous class of `super Earths' with radii between $1 - 4$\,R$_\oplus$, are super-sized rocky planets or mini gas giants. The best we can say is that planets with radius R $\gtrsim 1.5$\,R$_\oplus$ and for which we have both mass and radii measurements, commonly have mean densities consistent with a Neptune-like atmosphere \cite{Rogers2015}.  

These few cases where the bulk density is known are only mildly less confusing. Multiple possible compositions exist with wildly different surface prospects \cite{composition}. A higher density iron-enhanced rock with a thick hydrogen and helium atmosphere can give the same global density as a planet made predominantly from silicates. Similarly, a planet less dense than pure silicates could either have retained a thick atmosphere or be drowning under a global ocean. Changes in the stellar abundance could also produce alien rock compositions that may result in highly varied geologies. Like equilibrium temperature, the planet size is therefore a poor proxy for surface conditions. 

The full surface environment will be affected by a long list of properties that include magnetic fields, water delivery and retention, stellar activity, impact history, rotation, age, rock composition, tectonics and geochemical cycling \cite{NotEarth2}. The 2 - 3 quantities we can measure are only weakly linked to a fraction of these factors, making extrapolations to `habitability' almost meaningless.

\subsubsection*{The best of all possible worlds}

There are three metrics that are in common use for ranking exoplanets. Each employ a different method to combine observable properties into a numerical value. 

{\bf The Circumstellar Habitable Zone} (hereafter HZ) is the region around a star where liquid water could exist on the surface of an exact Earth-clone. In this region, the equilibrium temperature would give surface values between $0 - 100^{\circ}$C when combined with the Earth's greenhouse atmosphere \cite{HZ}. This region does not promise water will be present or that it can be maintained on any surface that differs from the Earth. But if a clone of our planet does exist, we will find it within the HZ.

{\bf The Habitability Index for Transiting Exoplanets} (HITE) has been recently developed to rank transiting exoplanets \cite{HITE}. Broadly speaking, the HITE weights planets within a modified HZ based on their radius and probable orbital eccentricity. The modification of the HZ shifts its location to allow for changes to the planet's gravity, but assumes atmospheric processes remain Earth-like. High HITE values correspond to planets orbiting centrally in the modified HZ with radii less than 1.5\,R$_\oplus$.

{\bf The Earth Similarity Index} (ESI) creates a ranking based on deviations from Earth values for properties derived from the observable quantities \cite{ESI}. This effectively compares Earth analogues with different sizes and equilibrium temperatures. Values between $0.8 - 1.0$ are said to be `Earth-like'. 

The goal of metrics for selecting targets in future habitability studies is to define a region of the observable parameter space that could include Earth-like worlds. However, the complex relationship between the observable properties and habitability is frequently ignored when interpreting the metric value. This leads to the false assumption that a higher ranking means a higher probability for life. 

That the surface degeneracies for the observable properties make this connection impossible is easily demonstrated within our Solar System. The similarity in size and equilibrium temperature of the Earth and Venus gives Venus an ESI value of 0.9 (assuming its true surface temperature is unknown as for all exoplanets). Yet even a spaceprobe cannot survive beyond two hours on the Venusian surface. The most wide-spread metric misuse is that the position of the HZ can be taken as a stand-in for the existence of life on a world. The only reason for such speculation is that the Earth has life and is within the Sun's HZ. Notably, so is the Moon. These false interpretations have led to the widespread belief that we have found `{\it Earth 2.0}', risking the same twenty year battle for funding that Martian programs faced in the wake of the controversial detection of life by the Viking mission.

\subsubsection*{What's in a number?}

Given the need for target selection (and possibly an innate desire to sort catalogues of objects) abandoning the use of metrics is not practical. But steps could be taken to minimise misrepresentation:

Firstly, any metric for target selection should go to zero when there is no possibility of spectroscopic data. This keeps the focus on the detectability of life. 

While changing naming conventions is never easy, the titles of the current metrics are particularly misleading. An improvement would reflect either the aim or the measured quantity of the metric. For example, referring to the HZ as the `Temperate Zone' places the emphasis correctly on stellar radiation. Similarly, `Detectable Environment Index' for HITE would clarify the observational purpose. While `similarity index' is a recognised term, more intuitive wording would be `Earth Scalability Index' to represent a value created by deviations from a given standard. A different tack would be to allocate a number based on factors against life, creating an `undetectable index' that is less easily related to alien neighbours.

The most important step is to discuss metrics as target selection tools and guard against over-reaching their applicability both in scientific literature and in material for a wider audience. Our knowledge is far from sufficient to comparatively rank the ability of planets to support life. Unless we want to risk destroying the chance to find out if the Earth is unique, we need to stop pretending that we already know.

\end{document}